\documentclass[10pt]{article}

\usepackage[margin=1in]{geometry}
\usepackage{graphicx}
\usepackage{amsmath,amssymb,braket}
\usepackage{cite}
\usepackage{authblk}
\usepackage{hyperref}
\usepackage{xcolor}
\usepackage{lineno}
\title{Saturable nonlinearities in a driven-dissipative bosonic quantum battery}

\author[1]{J. P. R. Leonel}
\author[1]{P. A. Brand\~ao\thanks{paulo.brandao@fis.ufal.br}}
\affil[1]{Instituto de F\'isica, Universidade Federal de Alagoas, Av. Lourival Melo Mota, S/N, Tabuleiro do Martins, 57072-970, Macei\'o, Alagoas, Brasil. }

\date{} % deixa vazio

\begin{document}

\maketitle

% ---------- Abstract ----------
\begin{abstract}
	We investigate the charging of a nonlinear quantum battery consisting of a single bosonic mode subject to a saturable nonlinearity, coherent driving, and dissipation. In contrast to Kerr-type anharmonicities, the saturable interaction induces a bounded and nonlinear distortion of the energy spectrum, leading to a progressive increase in the density of energy levels. We analyze the time evolution of the energy and ergotropy of the battery by solving a Lindblad master equation and show that the nonlinear spectral structure significantly affects both transient charging behavior and steady-state properties. Our results reveal that, for a broad range of parameters, the saturable nonlinearity enhances the maximum stored energy and modifies the ergotropy generation in the presence of losses. The interplay between dissipation and bounded spectral nonlinearity provides a controllable mechanism to tune energy storage and work extraction in bosonic quantum batteries.
\end{abstract}

% ---------- Seções ----------
\section{Introduction}

Quantum batteries are quantum mechanical devices designed to store energy and subsequently deliver it as useful work, potentially exhibiting quantum advantages over their classical counterparts through coherence, correlations, and collective quantum effects \cite{Alicki2013,Binder2015,Campaioli2017,Campaioli2024, ferraro2026opportunities}. The performance of a quantum battery is commonly characterized by figures of merit such as the stored energy, charging time, and charging power, together with the ergotropy, defined as the maximum extractable work under unitary and cyclic operations \cite{Allahverdyan2004,Francica2020}. These concepts have motivated a rapidly growing literature ranging from foundational bounds and protocols to platform-motivated proposals and open-system considerations \cite{Campaioli2017,Campaioli2024,Andolina2018,Andolina2019}.

A broad range of physical models has been explored, including interacting spins and many-body settings \cite{Le2018,Rossini2020}, cavity and light-matter based systems \cite{Farina2019,Crescente2020,Delmonte2021}, and autonomous or dissipative charging schemes \cite{Barra2019,Tabesh2020}. In realistic scenarios, environmental coupling is unavoidable and can strongly affect both transient charging and steady state behavior, making open system approaches central to assessing stability, charging performance, and extractable work \cite{Farina2019,Barra2019,Tabesh2020,Cakmak2020}. In this context, GKSL master equations provide a standard framework \cite{Gorini1976,Lindblad1976,Manzano2020}, enabling a quantitative description of driven-dissipative battery dynamics and steady states.

Continuous-variable (bosonic) quantum batteries are particularly appealing because they can in principle access large energy capacities, while allowing powerful analytical and numerical techniques inherited from quantum optics and open quantum systems \cite{GardinerZoller2004,Breuer2002,WallsMilburn2008}. A key direction in this class of models is spectral engineering through nonlinearities. Kerr-type responses are a paradigmatic mechanism: they distort the equally spaced oscillator ladder and can interpolate between harmonic and effectively few-level behavior as the nonlinearity increases \cite{Ukhtary2023}. This idea connects with the long-standing theory of driven-dissipative Kerr oscillators and optical bistability, where nonlinear spectral shifts lead to multistability, critical behavior, and strongly nonclassical states \cite{DrummondWalls1980,DrummondGardiner1980,Lugiato1984,Zhang2021,Kryuchkyan1996}. Kerr physics is also ubiquitous across platforms, including cavity QED and related nonlinear optical settings \cite{Imamoglu1997,Zhu2010,Gong2009}.

Despite its usefulness, Kerr nonlinearity produces an unbounded spectral distortion that grows with excitation number, which may render strongly anharmonic regimes overly rigid when large excitations are involved. This motivates considering bounded nonlinear mechanisms. A physically natural alternative is provided by saturable nonlinearities, which appear in nonlinear optical media, polaritonic and cavity QED settings, and in models of absorptive/dispersive bistability where the response saturates with intensity \cite{Lugiato1984,Imamoglu1997,Zhu2010, gatz1991soliton}. In such cases, the nonlinear spectral modification can be substantial at intermediate excitation while remaining asymptotically limited, offering a route to reshape the density of levels without an unbounded divergence.

Motivated by these ideas, in this work we investigate a driven-dissipative bosonic quantum battery endowed with a saturable nonlinearity. By solving the corresponding Lindblad master equation numerically (using standard open-system tools \cite{lambert2026qutip}), we characterize energy storage and work extraction through the stored energy and ergotropy, and analyze both transient dynamics and steady-state properties. We show that the saturable interaction produces a bounded spectral distortion associated with an effective densification of energy levels, which can enhance the accessible energy window and significantly modify ergotropy generation in the presence of losses. Our study complements Kerr-based quantum battery proposals \cite{Ukhtary2023} and connects quantum battery performance to broader driven-dissipative nonlinear oscillator physics \cite{DrummondWalls1980,Zhang2021} and recent continuous-variable battery protocols exploiting Gaussian resources and nonlinear driving/coupling \cite{Downing2023,Downing2024Hyperbolic,Downing2024Gaussian,Downing2025PRE}.

\section{Nonlinear quantum battery model}
\label{section2}

\subsection{General framework of a saturable nonlinear QB}
We consider a single-mode nonlinear cavity having frequency $\omega$ taken as a quantum battery. The charging protocol works as follows: At $t = 0^{-}$, the QB is in the initial state $\rho(0)$. During the time window $t\in[0,\tau]$, the QB interacts with an external driving field of frequency $\Omega$ and scalar amplitude $\alpha$. When all systems are interacting, the effective Hamiltonian has the form (setting $\hbar = 1$):
\begin{equation}\label{hamil1}
	\begin{split}
		h(t) &= h_b +\alpha(e^{i\Omega t}b + e^{-i\Omega t}b^{\dagger}) \\
		&=\omega b^{\dagger}b + \frac{\chi b^{\dag}b}{1 + n_s b^{\dagger}b} + \alpha(e^{i\Omega t}b + e^{-i\Omega t}b^{\dagger}),
	\end{split}
\end{equation}
where $b$ is the annihilation operator, $\omega$ is the angular frequency of the cavity, $\Omega$ is the angular frequency of the external drive, $\chi$ is the nonlinear parameter, $n_s$ is the saturable parameter, $\alpha$ is the scalar amplitude of the external driving and $h_b$ is the bare Hamiltonian of the QB.   This nonlinear effective model is motivated by the broad occurrence of saturable nonlinearities in nonlinear optics, optical bistability, photorefractive media and discrete nonlinear Schr\"odinger models \cite{drummond2014quantum, gatz1991soliton,man1997f, samuelsen2013statistical, aslan2011exact}. The term $\chi \hat n/(1+n_s\hat n)$, with $\hat n=b^\dagger b$,
can be regarded as a phenomenological Fock-space representation of an
intensity-dependent nonlinear response that saturates at large occupation
numbers. For small occupations, it gives a Kerr-like expansion, whereas
for large occupations the nonlinear energy shift becomes bounded. The nonlinear term can be formally expanded as $\chi b^{\dag}b/(1 + n_s b^{\dag}b) = \chi b^{\dag}b[1 - n_s b^{\dag}b + n_s^2(b^{\dag}b)^2 - ...]$. The  linear contribution induces a frequency shift $\omega \rightarrow \omega + \chi$ in the oscillation frequency of the QB. Keeping terms up to second-order in the expansion, the Hamiltonian of the QB takes the approximate form $(\omega + \chi)b^{\dag}b - n_s\chi b^{\dag}bb^{\dag}b$, which corresponds to a Kerr-type anharmonicity. The Kerr contribution has been recently considered in the context of quantum batteries \cite{Ukhtary2023}. 

\begin{figure}[]
	\includegraphics[width=0.8\linewidth]{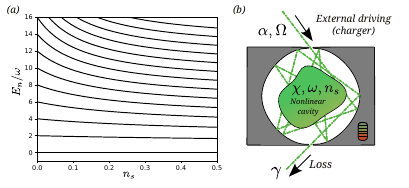}
	\centering
	\caption{  (a) Normalized eigenvalues $E_n/\omega = n + n(\chi/\omega) / (1 + nn_s)$ of the nonlinear QB as a function of $n_s$ for $\chi/\omega = 1$. (b) Sketch of the quantum battery system showing all the relevant parameters used in the model.   }
	\label{fig1}
\end{figure}

The eigenvalues $E_n$ of the Hamiltonian $h_b$ are given by 
\begin{equation}\label{nonlineareig}
	E_n = \omega n + \frac{n\chi }{1 + nn_s},
\end{equation}
so that, for fixed $\chi$, the eigenvalues saturate according to $E_n \sim \omega n + \chi/n_s$ as $n$ increases. This should be contrasted with the Kerr contribution which reads $E_n^{(\text{Kerr})} \sim (\omega + \chi)n - n_s\chi n^2$ and increases quadratically with $n_s$ and $\chi$. The   normalized eigenvalues $E_n/\omega$ are shown in Fig. \ref{fig1}(a).   Notice that for this approximation to make sense, one must have the parabolic term to be much smaller than the linear term, otherwise the energies would go to negative values.  Figure \ref{fig1}(a)   highlights the fact that for a fixed energy interval, say from   $E_n = 0$ to $E_n = 16$   as displayed in the picture, more energy levels become available as $n_s$ increases. Thus, the eigenvalues become more dense and this spectral rearrangement as $n_s$ increases induces significant changes in the performance of the QB, as is discussed below.

The Hamiltonian $H(t)$ in the rotating frame of the driver's frequency $\Omega$ can be obtained by defining the unitary operator $S \mathrel{:=} e^{i\Omega b^{\dag}b t}$ such that $H(t) = i\dot{S}(t)S^{\dag}(t) + S(t)h(t)S^{\dag}(t)$. In the case of $h(t)$ given by Eq. \eqref{hamil1}, $H(t) = H$ is time-independent and has the form
\begin{equation}\label{hamil2}
	H = \Delta b^{\dag}b + \frac{\chi b^{\dag}b}{1 + n_s b^{\dag}b} + \alpha(b + b^{\dag}),
\end{equation}
where $\Delta = \omega - \Omega$ is the detuning.   In the rotating frame, the diagonal part of the Hamiltonian, obtained by neglecting the driving term, is $H_r = \Delta b^{\dag}b + \chi b^{\dag}b/(1 + n_s b^{\dag}b)$. Its eigenvalues are $\varepsilon_n = \Delta n + \chi n / (1 + n_s n)$, with $n$ being a positive integer. The external driving term $\alpha(b+b^{\dag})$ induces transitions between the eigenstates $\ket{n} \leftrightarrow \ket{n+1}$ of $H_r$. Since the battery is nonlinear, the energy difference between these levels is not constant and can be characterized by 
\begin{equation}\label{spacingE}
	\delta_n(n_s,\chi) = \varepsilon_{n+1} - \varepsilon_n = \Delta + \frac{\chi}{(1 + n_s n)[1 + n_s(n+1)]}.
\end{equation}
Notice that in the linear case, $\delta_n(n_s,\chi = 0) = \Delta = \omega-\Omega$ and transitions between energy levels are more efficient if $\Delta \approx 0$, or $\omega \approx \Omega$. In the nonlinear case, the transitions are approximately resonant when the level spacing between connected states by the drive satisfies $\delta_n \approx 0$. This implies that the driving term (the charging) mixes these states more efficiently.

In the laboratory frame, the resonant condition can be stated as $E_{n+1} - E_n = \Omega$. Since $E_n$ is given by Eq. \eqref{nonlineareig}, we obtain $E_{n+1} - E_n = \omega + \chi/(1 + n_sn)[1 + n_s(n+1)]$ and substituting into $E_{n+1} - E_n - \Omega = 0$, one obtains $\delta_n(n_s,\chi) = 0$, where $\delta_n$ is exactly given by Eq. \eqref{spacingE}. Thus, in the nonlinear regime, energy is efficiently transferred between eigenstates if $\delta_n(n_s,\chi) = 0$, or
\begin{equation}\label{resonantcondition}
	\Delta + \frac{\chi}{(1 + n_s n)[1 + n_s(n+1)]} \approx 0.
\end{equation}
The above expression suggests that if $\chi > 0$, then one must have $\Delta < 0$ to balance the second term in the equation and vice-versa. In the following we assume that $\chi > 0$. Therefore, one should expect highly efficient energy transfers if the external drive frequency $\Omega$ is larger than the linear oscillator frequency $\omega$ $(\Delta = \omega - \Omega < 0)$.

We also assume that the QB is in contact with a large reservoir such that the density operator $\rho(t)$ in the rotating frame satisfies a Lindblad equation of the form $\dot{\rho}(t) = \mathcal{L}\rho(t)$, where
\begin{equation}\label{liou}
	\mathcal{L}[*] = -i[H,*] + \gamma \Big( b * b^{\dag} - \frac{1}{2}\Big\{ b^{\dag}b, * \Big\} \Big)
\end{equation}
is the Lindbladian superoperator and $\gamma$ the loss rate. We assume that dissipation acts during the charging dynamics and that the coupling between the nonlinear QB and the (zero temperature) bath is linear. The effect of $\gamma$, which tends to incoherently drive the QB to its ground state, compete with the nonlinear response of the QB and this interplay between dissipation and nonlinearity is the main focus of this paper.   A sketch of the nonlinear quantum battery is shown in Fig. \ref{fig1}(b). 

At $t = \tau$, the QB is disconnected from the charger and its energy $E(\tau)$ is calculated from the equation \begin{equation}
	E(\tau) = \text{tr} \left[  h_B \rho(\tau) \right],
\end{equation}
where $\text{tr}[*]$ is the trace operation. We also calculate the ergotropy $\mathcal{E}(\tau)$   and the mean charging power $P_E(\tau)$, defined by 
\begin{equation}
	\mathcal{E}(\tau) = \text{tr} \left[ h_B\rho(\tau) \right] - \text{tr}\left[ h_B\sigma(\tau)\right],
\end{equation}
\begin{equation}
	P_E(\tau) = \frac{E(\tau)}{\tau},
\end{equation} 
where $\sigma$ is the corresponding passive state of $\rho$. The passive state can be constructed directly from the spectral decomposition of $\rho = \sum_n \lambda_n \ket{\lambda_n}\bra{\lambda_n}$ and $h_B = \sum_{n}E_n \ket{E_n}\bra{E_n}$ written such that $E_{n+1} > E_{n}$ and $\lambda_{n+1} < \lambda_n$. It is defined by $\sigma = \sum_n \lambda_n \ket{E_n}\bra{E_n}$, which clearly commutes with the Hamiltonian $h_B$ used in the construction. The ergotropy gives a nonnegative real number representing the amount of energy in state $\rho$ that can be extracted as work under a unitary and cyclic process \cite{Allahverdyan2004}.

\subsection{Linear driven-dissipative quantum battery}

In this subsection, we briefly discuss the analytical solutions to the problem of charging a linear driven-dissipative quantum battery ($\chi = 0$), using the protocol discussed in the previous subsection. For $\chi = 0$, the Hamiltonian \eqref{hamil2} simplifies to $H_{\text{lin}} = \Delta b^{\dag}b + \alpha(b + b^{\dag})$ and the evolution of the mean value of the annihilation operator $\eta(t) = \langle b \rangle_t = \text{tr}[b\rho(t)]$, where $\rho$ satisfies the Lindblad equation with Liouvillian given by Eq. \eqref{liou}, is described by the ordinary linear differential equation
\begin{equation}\label{etaode}
	\dot{\eta} = -\left(  \frac{\gamma}{2} + i\Delta \right)\eta -i\alpha. 
\end{equation}
If the QB is initially in the vacuum state, $\rho(0) = \ket{0}\bra{0}$, we have $\eta(0) = \braket{0 | b | 0} = 0$ and the general solution of Eq. \eqref{etaode} turns out to be
\begin{equation}
	\eta(t) = -\frac{i\alpha}{\gamma/2 + i\Delta}\left[ 1 - e^{-(\gamma/2 + i\Delta)t} \right].
\end{equation}
At $t = \tau$, the energy $E_{\text{lin}}(\tau)$ and ergotropy $\mathcal{E}_{\text{lin}}(\tau)$ stored in the QB are given by
\begin{equation}\label{energylinear}
	E_{\text{lin}}(\tau) = \mathcal{E}_{\text{lin}}(\tau) = \omega \frac{\alpha^2}{(\gamma/2)^2 + \Delta^2}\Bigg[ 1 + e^{-\gamma \tau} - 2e^{-\gamma \tau /2} \cos(\Delta \tau) \Bigg].
\end{equation}
In the following sections, we compare these analytical solutions to the numerical ones obtained in the nonlinear $(\chi \neq 0)$ case. In all figures, energies are expressed in units of $\omega$, times in units
of $\omega^{-1}$, and rates/frequencies in units of $\omega$. Therefore, the dimensionless parameters used in the numerical simulations are
$\Delta/\omega$, $\chi/\omega$, $\alpha/\omega$, and $\gamma/\omega$,
while $n_s$ is dimensionless.

\section{\label{section3}Charging dynamics and steady states}

\begin{figure}[]
	\includegraphics[width=1\linewidth]{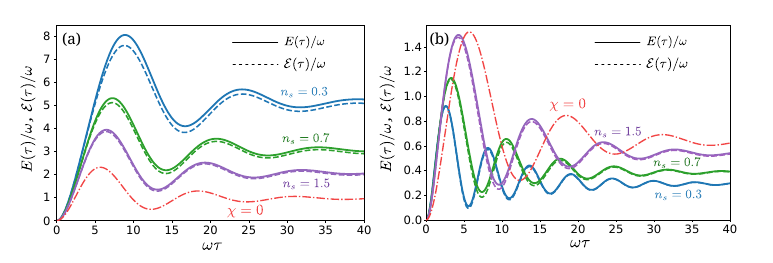}
	\centering
	\caption{  Normalized energy $E(\tau)/\omega$ (continuous lines) and ergotropy $\mathcal{E}(\tau)/\omega$ (dashed lines) of the nonlinear QB as a function of $\omega\tau$ for several values of the saturable parameter $n_s$ and for negative (a) $\Delta/\omega = -0.5$ and positive (b) $\Delta/\omega = 0.5$ detunings. The dotted-dashed red curves are the energy/ergotropy for the linear quantum battery [Eq. \eqref{energylinear}]. The other parameters used for this plot are given by $\chi/\omega = 1$, $\alpha/\omega = 0.5$ and $\gamma/\omega = 0.2$.  }
	\label{fig2}
\end{figure}

Let us start by considering what happens to the energy of the QB as soon as the interaction between the oscillator and external drive begins. In all cases we assume that $\rho(0) = \ket{0}\bra{0}$, where $\ket{0}$ is the ground state,  and numerically solve the Lindblad equation  by using the QuTiP package in Python   \cite{lambert2026qutip}. Figure~\ref{fig2} shows plots of the   normalized   energy and ergotropy as a function of $\omega\tau$ for several values of the saturable parameter $n_s$.   The plots show that, for negative detuning, finite saturable nonlinearities can enhance the stored energy and ergotropy relative to the corresponding linear benchmark. This behavior is not simply due to an increase in the density of levels, but rather to the fact that, for negative detuning, finite values of $n_s$ can bring selected neighboring transitions close to resonance with the external drive, as quantified by Eq.~\eqref{resonantcondition}.   This is in stark contrast to what was considered in the Kerr quantum battery as the (Kerr) nonlinearity increases (it behaves like a qubit) \cite{Ukhtary2023}. We point out that for large values of $n_s$, the QB behaves as a linear oscillator. The role of $n_s$ is thus to control the transition between two linear harmonic oscillator batteries, which differ by their density of states.

\begin{figure}[]
	\includegraphics[width=1\linewidth]{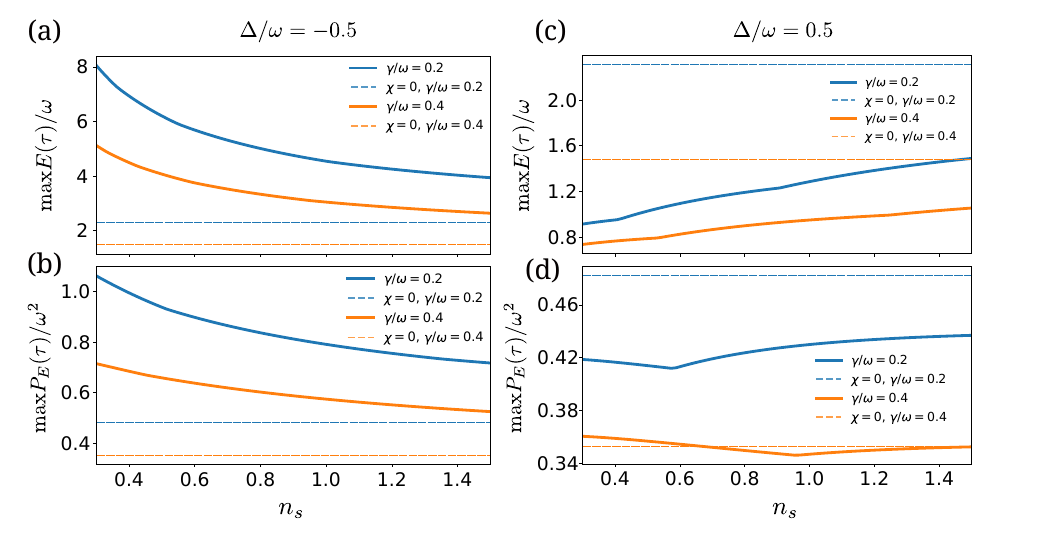}
	\centering
	\caption{  Maximum values of the normalized energy $E(\tau)/\omega$ and charging power $P_E(\tau)/\omega^2$ as a function of the saturable parameter $n_s$ for $\gamma = 0.2$ (blue) and $\gamma = 0.4$ (orange). (a-b) $\Delta < 0$ and (c-d) $\Delta > 0$. The dashed lines are the maximum values of energy and power obtained from the linear model [Eq. \eqref{energylinear}]. The other parameters are the same as in Fig. \ref{fig2} with the exception of $\omega\tau$ which takes the maximum value of 200 in this plot.}
	\label{fig3}
\end{figure}

Another aspect that Fig. \ref{fig2} reveals is that, for a range of values of the saturable parameter $n_s$ and fixed dissipation rate $\gamma$, the normalized energy/ergotropy exhibits a maximum at some time $\omega\tau$.   A plot of max$[E(\tau)/\omega]$ and max$[P_E(\tau)/\omega^2]$ as a function of $n_s$ for $\Delta < 0$ and $\Delta > 0$ cases is shown in Fig. \ref{fig3} for $\gamma/\omega = 0.2$ (same as in Fig. \ref{fig2}) and $\gamma/\omega = 0.4$. For this plot we take $\omega\tau = 200$. In parts (a) and (b) of Fig. \ref{fig3} one sees the energy and power for $\Delta < 0$, where we argued that eigenstates transitions are more efficient. This is reflected in these plots if one compares the continuous lines, representing the energy (a) and power (b), with the corresponding dashed lines which represent the maximum values of a linear QB. Although both the energy and power decrease with $n_s$, they remain higher than the corresponding linear limits. On the other hand, parts (c) and (d) suggest that the energy and power do not show any significant advantage over the linear QB, which is expected from the discussion of subsection 2.1.

It is usually very hard to find analytical and exact solutions to a nonlinear quantum model, and it is not the objective of this work to pursue in this direction, so that it is common to rely on numerical approaches to understand the charging dynamics. However, we can gain physical insight into the initial charging process, and try to grasp what are the most relevant parameters in this regime, by expanding the density operator $\rho(\tau)$ in a Taylor series in powers of $\tau$,
\begin{equation}\label{rhoexpansion}
	\rho(\tau) = \sum_{n=0}^{\infty} \frac{\tau^n}{n!}\rho^{(n)}(0) = \sum_{n=0}^{\infty} \frac{\tau^n}{n!} \mathcal{L}^{(n)}\rho(0),
\end{equation}
where $\rho^{(n)}(0)$ is the $n$th derivative of $\rho(\tau)$ at $\tau = 0$,
\begin{equation}
	\mathcal{L}^{(n)} = \overbrace{\mathcal{L}\mathcal{L}\mathcal{L}\cdots\mathcal{L}}^{n \text{ times}},
\end{equation}
and the Lindblad equation has been used to arrive at the second line equation in \eqref{rhoexpansion}. The linear and quadratic terms can be easily calculated and are given by
\begin{align}
	\mathcal{L}^{(1)}\rho(0) &= i\alpha\Big(\ket{0}\bra{1} - \ket{1}\bra{0}\Big), \label{L1} \\
	\mathcal{L}^{(2)}\rho(0) &=  2\alpha^2\ket{1}\bra{1} - \alpha^2\ket{0}\bra{0} - \alpha^2\sqrt{2}\Big( \ket{0}\bra{2} + \ket{2}\bra{0}  \Big) \notag \\
	&+ i\alpha\Bigg[ i\Bigg(  \Delta + \frac{\chi}{1 + n_s} \Bigg) -\frac{\gamma}{2} \Bigg]\ket{0}\bra{1} + i\alpha\Bigg[ i\Bigg(  \Delta + \frac{\chi}{1 + n_s} \Bigg) +\frac{\gamma}{2} \Bigg]\ket{1}\bra{0}. \label{L2}
\end{align}
Equation \ref{L1}, which is the first order contribution in $\tau$, shows that the external drive induces the creation of coherences between the ground state $\ket{0}$ and the first excited state $\ket{1}$. The nonlinear term in the Hamiltonian begins contributing to the charging dynamics only at second order in $\tau$, as seen in Eq. \ref{L2}. Also, it impacts not only the energy $E(\tau)$ of the QB but the coherence, again between the fundamental and first excited states. At this order, the external drive has already induced coherences between $\ket{0}$ and $\ket{2}$ and the dissipative parameter $\gamma$ competes with the detuning $\Delta$ and the nonlinear parameters $\chi$ and $n_s$ in the generation of coherences. Although cumbersome, this approach is straightforward to apply and can generate Taylor series expansions for several quantities of interest, such as the energy. Pad\'e approximants can then be used to obtain analytical approximations \cite{bender1999advanced}. We do not follow this approach here however and rely on numerical approximations.  

\begin{figure}[]
	\includegraphics[width=0.8\linewidth]{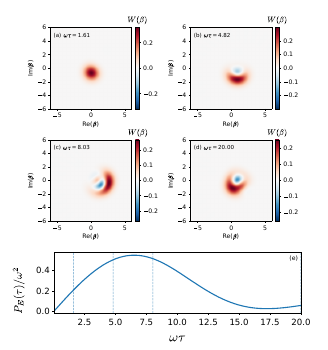}
	\centering
	\caption{  (a-d) Wigner function $W(\beta)$ during charging for several times and (e) normalized power as a function of $\omega \tau$. The vertical lines mark the respective times for the Wigner functions. For this plot we take $n_s = 1$, $\alpha/\omega = 0.3$ and $\gamma/\omega = 0.01$ and $\Delta/\omega = -0.5$. }
	\label{fig4}
\end{figure}

To show that the presence of a nonlinear response induces nonclassicality in the state of the QB, during its charging, Fig. \ref{fig4} shows the evolution of the Wigner function $W(\beta)$ from the beginning of the interaction until $\tau = 20$ for some selected set of parameters. We should expect a classical state with a positive Wigner function in the limits where $\alpha \gg \gamma$ or $\gamma \gg \alpha$. Negative values of the Wigner function are clearly present in the state and eventually disappear once the dissipative interaction becomes important. The presence of negative values in the Wigner function can only be explained by the onset of nonlinearities since it is well known that a driven linear oscillator evolves under the form of a coherent state, which is described by a nonnegative Wigner function \cite{drummond2014quantum}.   We emphasize that the present work does not claim a general quantum advantage over all possible classical analogues. Establishing such an advantage would require a resource-constrained comparison with a suitably defined classical nonlinear oscillator. What our results show is that the saturable nonlinear quantum model can generate nonclassical states, as witnessed by the negativity of the Wigner function, and that the nonlinear spectral structure modifies energy storage, ergotropy, and charging power relative to the corresponding linear quantum benchmark.

%(delta=.1, chi=1, n_s=1.0, alpha=0.3, tf=20, gamma=0.01)

Finally, we investigate the steady states of the QB in the presence of nonlinearity and dissipation. The solution to the Liouville equation $\dot{\rho} = \mathcal{L}\rho$, in the case where the superoperator $\mathcal{L}$ is diagonalizable, has the form
\begin{equation}
	\rho(\tau) = \sum_{n} c_n(\tau)\rho_n,
\end{equation}
where $c_n(\tau) = \exp(\lambda_n \tau )\text{tr}[r_n\rho(0)]$ \cite{minganti2019quantum}. The eigenvector operators $\rho_n$ and $r_n$ satisfy $\mathcal{L}\rho_n = \lambda_n \rho_n$ and $\mathcal{L}^{\dag} r_n = \lambda_n^* r_n$ with $\lambda_n$ and $\lambda_n^*$ the corresponding eigenvalues. It can be demonstrated that for any Lindbladian superoperator, $\text{Re}(\lambda_n) \leq 0$ and that $\lambda_{0} = 0$ is always an eigenvalue. The spectrum can thus be ordered in such a way that $|\text{Re}(\lambda_{0})| < |\text{Re}(\lambda_1)| < |\text{Re}(\lambda_2)| < ... < |\text{Re}(\lambda_n)|$. The steady state $\rho_{ss}$ is thus related to the eigenvector operator $\rho_0$ corresponding to the eigenvalue $\lambda_{0} = 0$ which satisfies $\dot{\rho}_{ss} = 0$. In practical terms, it is important to understand the structure of the steady states since one may not be able to precisely control the charging time in an experiment, which would probably increase the number of resources needed. In this case, it is best to let the system evolve until the contributions of the eigenstates $\rho_n$ for $n \geq 1 $ to die out, leaving only the steady state $\rho_{ss} = \rho_0/\text{tr}\rho_0$ relevant to the final dynamics.

\begin{figure}[hbt]
	\includegraphics[width=0.6\linewidth]{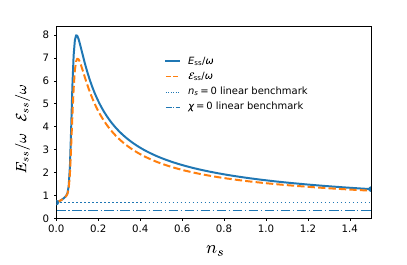}
	\centering
	\caption{  Normalized energy $E_{ss}/\omega$ and ergotropy $\mathcal{E}_{ss}/\omega$ of the steady state $\rho_{ss}$ as a function of $n_s$. The horizontal dashed lines indicate the analytical steady-state energies of the two linear benchmarks: the $n_s=0$ shifted linear oscillator and the $\chi=0$ ordinary linear oscillator. Parameters: $\Delta/\omega = -0.5$, $\chi/\omega  = 1$, $\alpha/\omega = 0.3$ and $\gamma/\omega = 0.2$. } 
	\label{fig5}
\end{figure}

Figure~5 shows the steady-state energy $E_{\rm ss}/\omega$ and
ergotropy $\mathcal{E}_{\rm ss}/\omega$ as functions of the saturation
parameter $n_s$. The horizontal dotted and dot-dashed lines denote two analytically solvable linear benchmarks. The dotted line corresponds to the $n_s=0$
limit, for which the saturable term reduces to $\chi \hat n$ and the
battery behaves as a shifted linear oscillator with frequency
$\omega+\chi$ and effective detuning $\Delta+\chi$. Its steady-state
energy is
\begin{equation}
	E_{\rm ss}^{(n_s=0)}
	=
	\frac{(\omega+\chi)\alpha^2}
	{\left(\gamma/2\right)^2+\left(\Delta+\chi\right)^2}.
\end{equation}
The dot-dashed line corresponds to the ordinary linear oscillator obtained for
$\chi=0$ ($n_s \rightarrow \infty$),
\begin{equation}
	E_{\rm ss}^{(\chi=0)}
	=
	\frac{\omega\alpha^2}
	{\left(\gamma/2\right)^2+\Delta^2}.
\end{equation}
The curves show that the saturable nonlinearity induces a rapid increase
of both $E_{\rm ss}$ and $\mathcal{E}_{\rm ss}$ as $n_s$ is increased from
zero. For finite values of $n_s$ considered in this plot, the steady-state energy and ergotropy exceed both linear benchmarks, indicating a finite-saturation enhancement associated with the nonlinear reshaping of the spectrum. This enhancement is not an asymptotic large-$n_s$ effect: in the limit $n_s\to\infty$, the saturable contribution
$\chi\hat n/(1+n_s\hat n)$ vanishes and the model approaches the
ordinary linear oscillator. Accordingly, the steady-state energy returns
to the $\chi=0$ benchmark for sufficiently large $n_s$.

\section{\label{section5}Conclusions}

 In this work, we investigated the charging dynamics of a driven-dissipative bosonic quantum battery subject to a saturable nonlinearity. The saturable interaction produces a bounded deformation of the harmonic spectrum, in contrast to Kerr-type nonlinearities, whose spectral distortion grows unboundedly with excitation number. This bounded nonlinear response allows one to continuously tune the level structure of the battery between two linear limiting regimes. A central result of our analysis is that the performance enhancement is strongly detuning dependent. By examining the diagonal part of the Hamiltonian in the rotating frame, we showed that the effective detuning between neighboring Fock states is
\begin{equation}
	\delta_n(n_s,\chi)
	=
	\Delta+
	\frac{\chi}
	{(1+n_s n)[1+n_s(n+1)]}.
\end{equation}
Since the coherent drive couples neighboring states, efficient charging is expected when $\delta_n(n_s,\chi)$ becomes small for the transitions participating in the dynamics. For $\chi>0$, this mechanism is most clearly activated for negative detuning, $\Delta<0$, where finite values of $n_s$ can compensate the bare detuning and bring selected transitions close to resonance with the external drive.

This resonance-assisted spectral engineering provides a clear physical interpretation of the numerical results. In the negative-detuning regime, the saturable nonlinear battery can outperform the corresponding linear benchmark in both maximum stored energy and maximum average charging power. In contrast, for positive detuning, such an enhancement is strongly suppressed, as the nonlinear contribution shifts the relevant transitions further away from resonance. Thus, the advantage of the saturable response is not universal, but parameter dependent: it emerges when the nonlinear spectral deformation improves the resonance conditions for the charging process. We also analyzed the ergotropy dynamics and found that a significant fraction of the stored energy remains extractable as work during the transient charging process. The phase-space analysis further showed that the nonlinear dynamics can generate states with negative Wigner functions, demonstrating the emergence of nonclassicality during charging. We stress, however, that this nonclassicality should not be interpreted as a general proof of quantum advantage over all possible classical analogues. Such a claim would require a separate resource-constrained comparison with a suitably defined classical nonlinear oscillator. Finally, the steady-state analysis revealed that finite saturable nonlinearities can also enhance the asymptotic stored energy and ergotropy relative to the linear limiting benchmarks in the detuned regime considered. This enhancement is not an asymptotic large-$n_s$ effect: in the limit $n_s\to\infty$, the saturable contribution vanishes and the ordinary linear oscillator is recovered. Instead, the improvement occurs in a finite range of $n_s$, where the nonlinear spectral reshaping remains appreciable.

Overall, our results indicate that saturable nonlinearities provide a viable and physically motivated mechanism to engineer spectral properties and control energy storage in open continuous-variable quantum batteries. The present analysis contributes to the broader understanding of how nonlinear spectral engineering can be used to tailor charging performance and work extractability in realistic driven-dissipative quantum systems.

\section*{Declarations}

\subsection*{Funding}
JPRL acknowledges the financial support of FAPEAL and PAB acknowledges the financial support of CNPq.

\subsection*{Competing interests}
The authors declare that they have no competing interests.

\subsection*{Data and code availability}
The Python/QuTiP scripts used to generate the numerical results and figures
reported in this work are publicly available at \url{https://github.com/paulocabf/SaturableQuantumBattery}. The numerical data supporting
the plots can be reproduced from the scripts provided in the repository.

% ---------- Bibliografia ----------
\bibliographystyle{unsrt}
\bibliography{references}

\end{document}